\let\accentvec\vec
\documentclass[runningheads,a4paper]{llncs}

\let\vec\accentvec

\usepackage{amssymb}
\usepackage{graphicx}

\usepackage{amsmath}
\usepackage{amsmath,bm}
\usepackage{url}
\urldef{\mailsa}\path|{alfred.hofmann, ursula.barth, ingrid.haas, frank.holzwarth,|
\urldef{\mailsb}\path|anna.kramer, leonie.kunz, christine.reiss, nicole.sator,|
\urldef{\mailsc}\path|erika.siebert-cole, peter.strasser, lncs}@springer.com|
\newcommand{\keywords}[1]{\par\addvspace\baselineskip
\noindent\keywordname\enspace\ignorespaces#1}

\usepackage{amsmath}

\graphicspath{{./fig/}}

\begin{document}

\mainmatter  


\title{Evidential communities for complex networks}

\titlerunning{Evidential communities for complex networks}

%
%

\author{Kuang Zhou\inst{1,2} \and Arnaud Martin\inst{2}
 \and Quan Pan\inst{1}}
\authorrunning{Kuang Zhou et al.} 
%
%
\institute{School of Automation, Northwestern Polytechnical University, \\ Xi'an, Shaanxi 710072, PR China
\and
IRISA, University of Rennes 1, Rue E. Branly, 22300 Lannion, France
\\
kzhoumath@163.com,  Arnaud.Martin@univ-rennes1.fr, quanpan@nwpu.edu.cn}

\toctitle{Lecture Notes in Computer Science}
\tocauthor{Authors' Instructions}
\maketitle

\begin{abstract}
Community detection is of great importance for understanding graph structure in social networks. The communities in real-world networks are often overlapped, {\em i.e.} some nodes may be a member of multiple clusters. How to uncover the overlapping communities/clusters in a complex network is a general problem in data mining of
network data sets. In this paper, a novel algorithm to identify overlapping communities in complex networks by
a combination of an evidential modularity function, a spectral mapping method and evidential $c$-means clustering is devised. Experimental results indicate that this detection approach can take advantage of the theory of belief functions, and preforms good both at detecting  community structure and determining the appropriate number of clusters. Moreover, the credal partition obtained by the proposed method could give us a  deeper insight into the graph structure.
\end{abstract}

\section{Introduction}
In order to have a better understanding of organizations and functions in the real networked system, the community structure, or the clustering in the graph is a primary feature that should be taken into consideration~\cite{fortunato2010community}. As a result,
community detection, which can extract specific structures from
complex networks,  has attracted considerable attention crossing many areas from physics, biology, and economics to sociology~\cite{costa2011analyzing}, where systems are often represented as graphs.

Generally, a community in a network is a subgraph whose nodes are densely connected within itself but sparsely connected with the rest of the network~\cite{zhang2007identification}. Many of the community detection approaches are in the frame of probability theory, that is to say, one actor in the network can belong to only one community of the graph~\cite{newman2004fast,girvan2002community}. However, in real-world networks, each node can fully or partially  belong to more than one associated community, and thus communities often overlap to some extent~\cite{palla2005uncovering,wang2009adjusting}. For instance, in collaboration networks, a researcher may be active in many areas but with different levels of commitment, and in social networks, an actor usually has
connections to several social groups like family, friends, and colleagues. In  biological networks, a node might have
multiple functions~\cite{palla2005uncovering}.

In the last decades, for identifying such clusters that are not necessarily disjoint, there is growing interest in overlapping community detection algorithms. Zhang et al.~\cite{zhang2007identification} devised a novel algorithm to identify overlapping communities in complex networks based on fuzzy $c$-means (FCM). Nepusz et al.~\cite{nepusz2008fuzzy} created an optimization
algorithm for determining the optimal fuzzy membership degrees, and a new fuzzified variant of the modularity function is introduced to determine the number of communities. Havens et al.~\cite{havens2013soft,havens2013clustering} discussed a new formulation of a fuzzy validity index and pointed out this modularity measure performs better compared with the existing ones.

As can be seen, most of methods for uncovering the overlapping community structure are based on the idea of fuzzy partition, which subsumes crisp partition, resulting in greater expressive power of fuzzy community detection compared with hard ones. Whereas credal partition~\cite{denoeux2004evclus}, which is even more general and allows in some cases to gain deeper insight into the structure of the data, it has not been applied to community detection.

In this paper, an algorithm for detecting overlapping community structure is proposed based on credal partition. An evidential modular function is introduced to determine the optimal number of communities. Spectral relaxation and evidential $c$-means are conducted to obtain the basic belief assignment (bba) of each nodes in the network. The experiments on two well-studied networks show that meaningful partitions of the graph could be obtained by the proposed detection approach and it indeed could provide us more informative information of the graph structure than the existing methods.

\section{Background}

\subsection{Modularity-based community detection}
Let $G(V,E,W)$ be an undirected network, $V$ is the set of  $n$ nodes, $E$ is the set of $m$ edges, and $W$ is a $n \times n$ edge weight matrix with elements $w_{ij}, i,j=1,2,\cdots,n$. The objective of the hard (crisp) community detection is to divide graph $G$ into $c$ clusters, denoted by \begin{equation}
\Omega=\{\omega_1,\omega_2,\cdots,\omega_c\},
\label{DF}
\end{equation}
and  each node should belong to one and only one of the detected communities~\cite{nepusz2008fuzzy}. Parameter $c$ can be given in advanced or determined by the detection method itself.

The modularity, which measures  the quality of a partition of a graph, was first introduced by
Newman and Girvan~\cite{newman2004finding}. This validity index measures how good a specific community structure is by calculating the difference between the actual edge density intra-clusters in the obtained partition and the expected one under some null models, such as random graph. One of the most popular form of modularity is given by~\cite{fortunato2010community}. Given a partition with $c$ group shown in Eq.~(\ref{DF}),  and let $\left \| W \right \|=\sum_{i,j=1}^n w_{ij}$, $k_i=\sum_{j=1}^n w_{ij}$, its modularity can be defined as:
\begin{equation}\label{hard_modu}
  Q_h=\frac{1}{\left \| W \right \|}\sum_{k=1}^c \sum_{i,j=1}^n (w_{ij}-\frac{k_i k_j}{\left \| W \right \|})\delta_{ik} \delta_{jk},
\end{equation}
where $\delta_{ik}$ is one if vertex $i$ belongs to the $k_{th}$ community, 0 otherwise.

The communities of graph $G$ can be detected by modularity optimization, like spectral clustering algorithm~\cite{smyth2005spectral}, which aims at finding the optimal partition with the maximum modularity value~\cite{fortunato2010community}.

\subsection{Belief function theory and evidential $c$-means}
The credal partition, a general extension of the crisp and fuzzy ones in the theoretical framework of belief function theory, has been introduced in~\cite{denoeux2004evclus,masson2008ecm}. Suppose the discernment frame of the clusters is $\Omega$ as in Eq.~\eqref{DF}. Partial knowledge regarding the actual cluster node $n_i$ belongs to can be represented by a basis belief assignment defined as a function $m$ from the  power set of $\Omega$ to $[0,1]$, verifying $\sum_{A \subseteq \Omega} m(A)=1$. Every $A\in 2^\Omega$ such that $m(A)>0$ is called a focal element. The credibility and plausibility functions are defined in Eq.~\eqref{bel} and Eq.~\eqref{pl}.
  \begin{equation}
  Bel\text{(}A\text{)}=\sum_{\emptyset \neq B\subseteq A} m\text{(}B\text{)}, \forall A\subseteq \Omega,
  \label{bel}
  \end{equation}
   \begin{equation}
   Pl\text{(}A\text{)}=\sum_{B\cap A \neq \emptyset}  m\text{(}B\text{)}, \forall A\subseteq \Omega.
   \label{pl}
   \end{equation}

Each quantity $Bel(A)$ represents the degree to which the evidence supports $A$, while $Pl(A)$ can be interpreted as an upper bound on the degree of support that could be assigned to $A$ if more specific information is  available~\cite{TBM}. The function $pl:\Omega \rightarrow [0,1]$ such that $pl(\omega) = Pl(\{\omega\})$ is called the contour function associated to $m$.

The bbas in the credal level can be expressed in the form of probabilities by  pignistic transformation~\cite{denoeux2004evclus}, which is defined as
\begin{equation}
 BetP(\omega_i)=\sum_{\omega_i  \in A \subseteq \Omega } \frac{m(A)}{|A|(1-m(\emptyset))},
 \label{pignistic}
\end{equation}
where $|A|$ is the number of elements of $\Omega$ in $A$.

Evidential $c$-means (ECM)~\cite{masson2008ecm} is a direct generalization of FCM. The optimal credal partition is obtained by  minimizing the following objective function:
\begin{equation}
J_{\mathrm{ECM}}=\sum\limits_{i=1}^{n}\sum\limits_{A_j\subseteq \Omega,A_j \neq \emptyset}|A_j|^\alpha m_{i}(A_j)^{\beta}d_{ij}^2+\sum\limits_{i=1}^{n}\delta^2m_{i}(\emptyset)^{\beta},
\label{JECM}
\end{equation}
\noindent constrained on
\begin{equation}
\sum\limits_{A_j\subseteq \Omega,A_j \neq \emptyset}m_{i}(A_j)+m_{i}(\emptyset)=1,
\label{ECMconstraint}
\end{equation}
where $m_{i}(A_j)$ is the bba of $n_i$ given to the nonempty set $A_j$, while $m_{i}(\emptyset)$ is the bba of $n_i$ assigned to the emptyset.
The value $d_{ij}$ denotes the distance between $n_i$ and the barycenter associated to $A_j$, and $|\cdot|$ is the cardinal of the set. Parameters $\alpha,\beta,\delta$ are adjustable and can be determined based on the requirement.

\section{Evidential community detection}
Before presenting the credal partition of a graph $G(V,E,W)$, the hard and fuzzy partitions are firstly recalled. The crisp partition can be represented by a matrix $U^h=(u_{ik})_{n \times c}$, where $u^h_{ik}=1$ if the $i_{th}$ node $n_i$ belongs to the $k_{th}$ cluster $\omega_i$ in the partition, and $u^h_{ik}=0$ otherwise. From the property of this partition, it clearly should satisfy that $\sum_{k=1}^c u^h_{ik}=1, i=1,2,\cdots,n$. The generalization of the hard partition, following that a node may belong to more communities  than one but with different degrees, can be
described by the fuzzy partition matrix $U^f=(u_{ik})_{n \times c}$, where $u^f_{ik}$ is not restricted in $\{0,1\}$ but can attain any real value from the interval $[0,1]$. The value $u^f_{ik}$ could be interpreted as a degree of membership of $n_i$ to community $\omega_k$.

The credal partition of $G$, which refers to the framework of belief function theory, can be represented by  a $n$-tuple: $M=(\bm{m}_1,\bm{m}_2,\cdots,\bm{m}_n)$. Each \linebreak $\bm{m}_i=\{m_{i1},m_{i2},\cdots,m_{i2^c}\}$ is  a bba in a $2^c$-dimensional space, where $c$ is the  cardinality  of the
given discernment frame of communities $\Omega=\{\omega_1,\omega_2,\cdots,\omega_c\}$ as  before, and $\omega_i$ denotes the $i_{th}$ detected community. Note that $\Omega$ is the discernment frame in the framework of belief function theory.

\subsection{The evidential modular function}

Similar to the fuzzy modularity by Nepusz et al.~\cite{nepusz2008fuzzy} and by  Havens et al.~\cite{havens2013soft}, here we introduce an evidential modularity:

\begin{equation}\label{evi_modu}
  Q_e=\frac{1}{\left \| W \right \|}\sum_{k=1}^c \sum_{i,j=1}^n (w_{ij}-\frac{k_i k_j}{\left \| W \right \|})pl_{ik} pl_{jk},
\end{equation}
where $\bm{pl}_{i}=\{pl_{i1},pl_{i2},\cdots,pl_{ic}\}$ is the contour function associated to $m_i$, which describes the upper value of our belief to the proposition that the $i_{th}$ node belongs to the $k_{th}$ community.

Let $\bm{k}=(k_1,k_2,\cdots,k_n)^{T}$, $B=W-\bm{k}^T\bm{k}/\left \| W \right \|$, and $PL=(pl_{ik})_{n\times c}$, then Eq.~\eqref{evi_modu} can be rewritten as:

\begin{equation}\label{evi_modu1}
  Q_e=\frac{\mathrm{trace}(PL~B~PL^T)}{\left \| W \right \|}.
\end{equation}
$Q_e$ is a directly extension of the crisp  modularity function~\eqref{hard_modu}. When the credal partition degrades into the hard one, $Q_e$ is equal to $Q_h$.
\subsection{Spectral mapping}

White and Smyth~\cite{smyth2005spectral} showed that optimizing the modularity measure $Q$  can be reformulated as a spectral relaxation problem and proposed spectral clustering algorithms that seek to maximize $Q$. By
eigendecomposing a related matrix, these methods can map graph data points into Euclidean space, the clustering problem on which space is of equivalence to that on the original graph.

Let $A=(a_{ij})_{n \times n}$ be the adjacent matrix of the graph $G$. The adjacency matrix for a weighted
graph is given by the matrix whose element $a_{ij}$ represents the weight $w_{ij}$ connecting nodes $i$ and $j$. The degree matrix $D=(d_{ii})$ is the diagonal matrix whose elements are the degrees of the nodes of $G$, {\em i.e.}  \linebreak $d_{ii}=\sum_{j=1}^n a_{ij}$. The eigenvectors of the transition matrix $\mathcal{M}=D^{-1} A$ are used.

Verma and Meila~\cite{verma2003comparison} and Zhang et al.~\cite{zhang2007identification} suggested to use the eigenvectors of a generalised eigensystem $Ax=\lambda Dx$, and pointed out that it is mathematically equivalent and  numerically more stable than computing the eigenvectors of  matrix $\mathcal{M}$~\cite{verma2003comparison}. To partition the nodes of the graph into $c$ communities, the top $c-1$  eigenvectors of
the above eigensystem are used to map the graph data into points in the Euclidean space, where the traditional clustering methods, such as $c$-means (CM), FCM and ECM can be evoked.
\subsection{Evidential community detection scheme}
Let $C$ be the upper bound of the number of communities. The evidential community detection scheme is displayed as follows:
\begin{description}
  \item[S.1] Spectral mapping:\\ For $2 \leq c \leq C$, Find the top $c$  generalized  eigenvectors $E_c=[e_1,e_2,\cdots,e_c]$ of the eigensystem $Ax=\lambda Dx$, where $A$ and $D$ are the adjacent and the degree matrix respectively.
  \item[S.2] Evidential $c$-means:\\ For each value of $c$ ($2\leq c \leq C$), let $E_c=[e_2,\cdots,e_c]$. Use  ECM to partition the $n$ samples (each row of $E_c$ is a sample data on the $c-1$ dimensional Euclidean space) into $c$ classes. And we can get a credal partition $M$ for the graph.
  \item[S.3] Choosing the number of communities:\\ Find the suitable number of clusters and the corresponding evidential partition scheme by maximizing the evidential modular function $Q_e$.
\end{description}

In the algorithm, $C$ can be determined by the original graph. It is an empirical range of the community number of the network. If $c$ is given, we can get a credal partition using the proposed method and then the evidential modularity can be derived. The modularity is a function of $c$ and it should peak around the optimum value of $c$ for the given network. As in ECM, the number of parameters to be optimized is exponential in the number of communities and linear in the number of nodes. When the number of communities is large, we can reduce the complexity by considering only a subclass of bbas with a limited number of focal sets~\cite{masson2008ecm}.
\section{Experimental results}
To evaluate the proposed method in this paper, two real-world networks  are discussed in this section. A comparison for the detected communities by credal, hard and fuzzy partitions is also illustrated to show the advantages of evidential community structure over others.
\subsection{Zachary's Karate Club}
\begin{center}
\begin{figure}[!thbt]
\centering
	\includegraphics[width=0.45\linewidth]{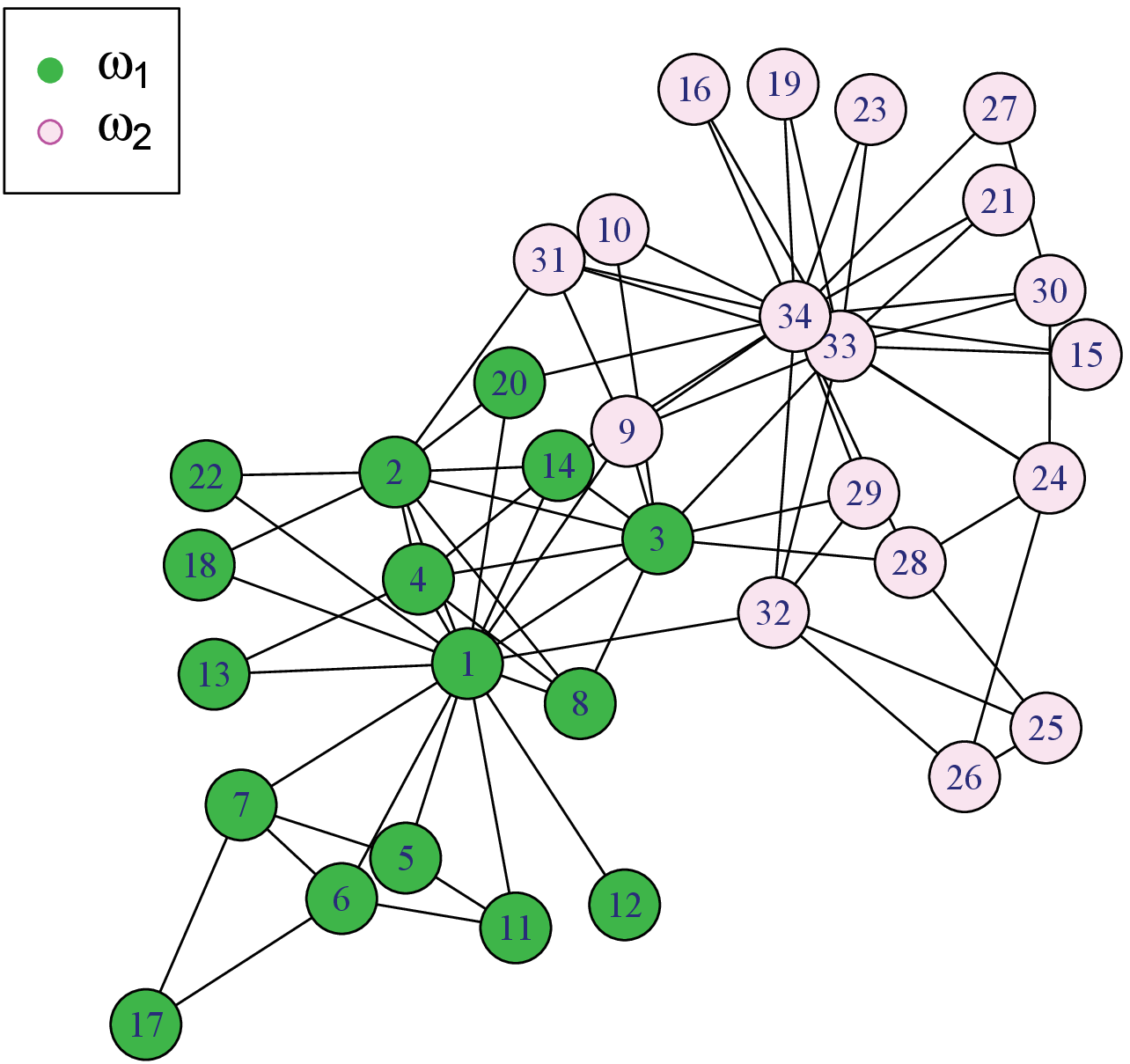}
	\hfill
	\includegraphics[width=.45\linewidth]{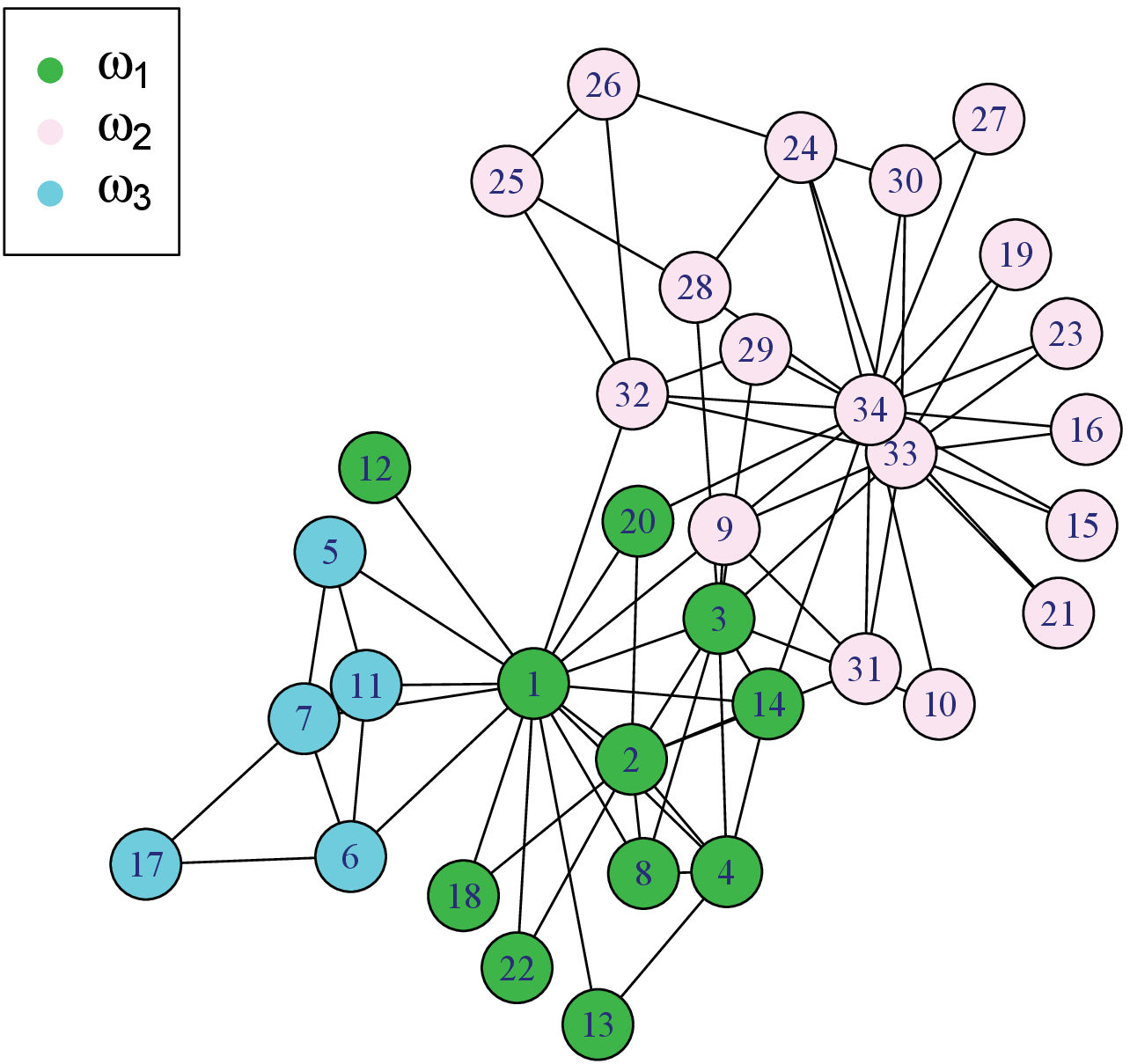}
	\hfill
    \parbox{.45\linewidth}{\centering\small a. Original network}
	\hfill
	\parbox{.45\linewidth}{\centering\small b. CM}
	\hfill
   \includegraphics[width=0.45\linewidth]{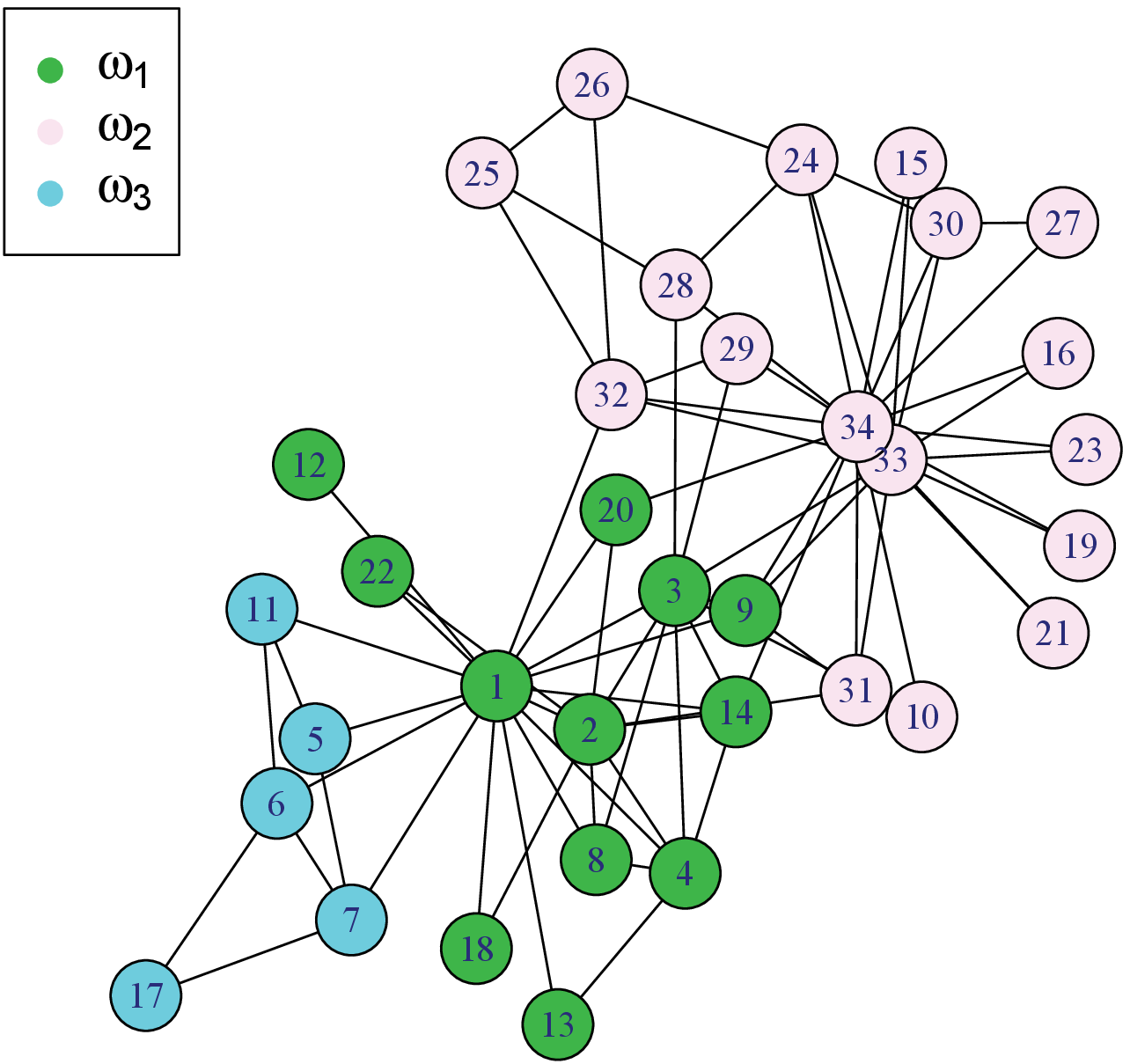}
	\hfill
	\includegraphics[width=0.45\linewidth]{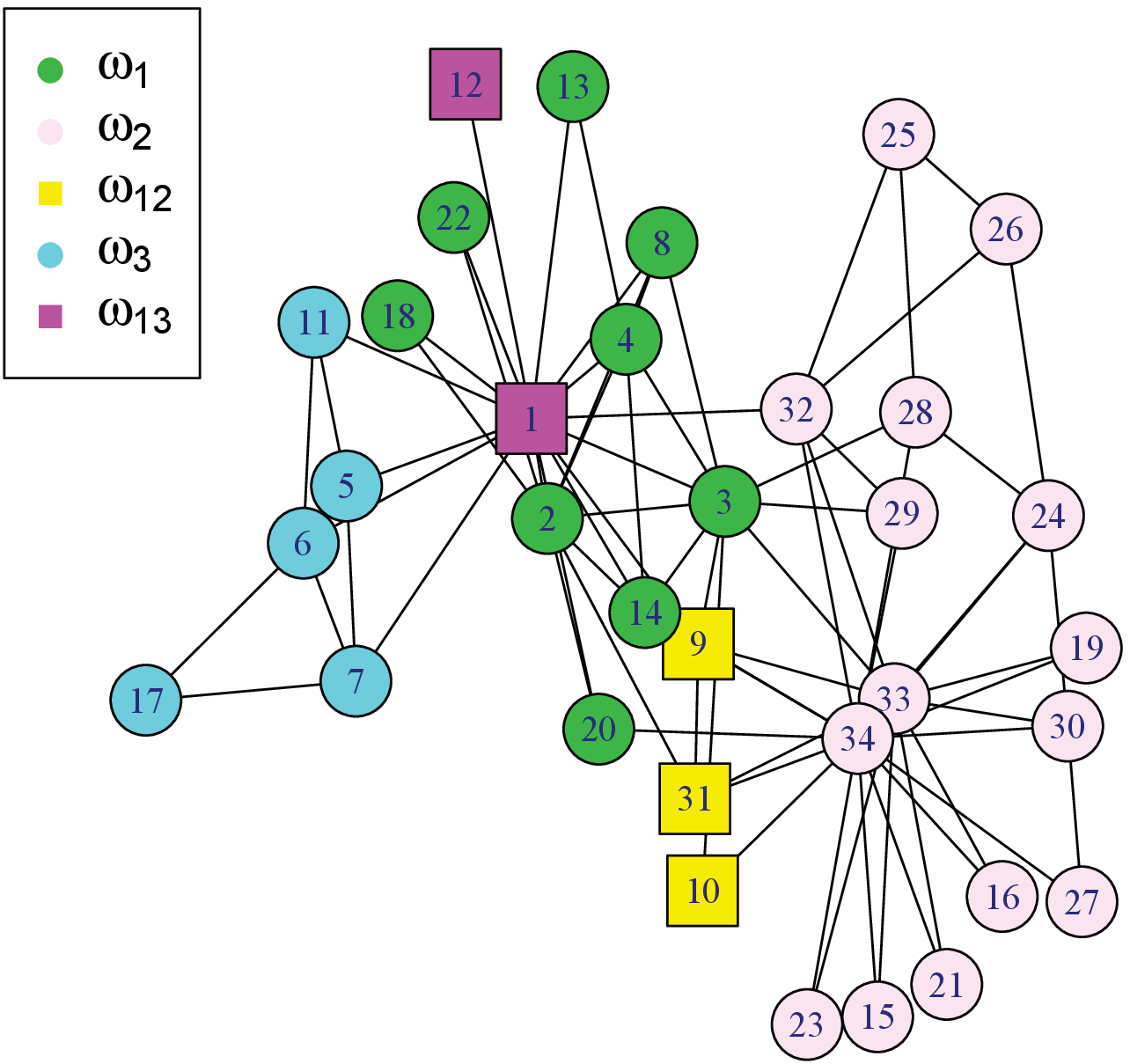}
	\hfill
    \parbox{.45\linewidth}{\centering\small c. FCM}
	\hfill
	\parbox{.45\linewidth}{\centering\small d. ECM}
	\hfill

     \caption{Karate Club.}
     \label{karate}
\end{figure}
\end{center}

The Zachary's Karate Club~\cite{zachary1977information} is an undirected  graph which consists of 34 vertices and 78 edges, describing the friendship between  members of the club  observed by Zachary in his
 two-year study.  This club is visually divided into two parts, due to an incipient conflict
between the president and  instructor (see  Fig.~\ref{karate}-a).

\begin{center}
\begin{figure}[!thbt]
\centering
	\includegraphics[width=0.45\linewidth]{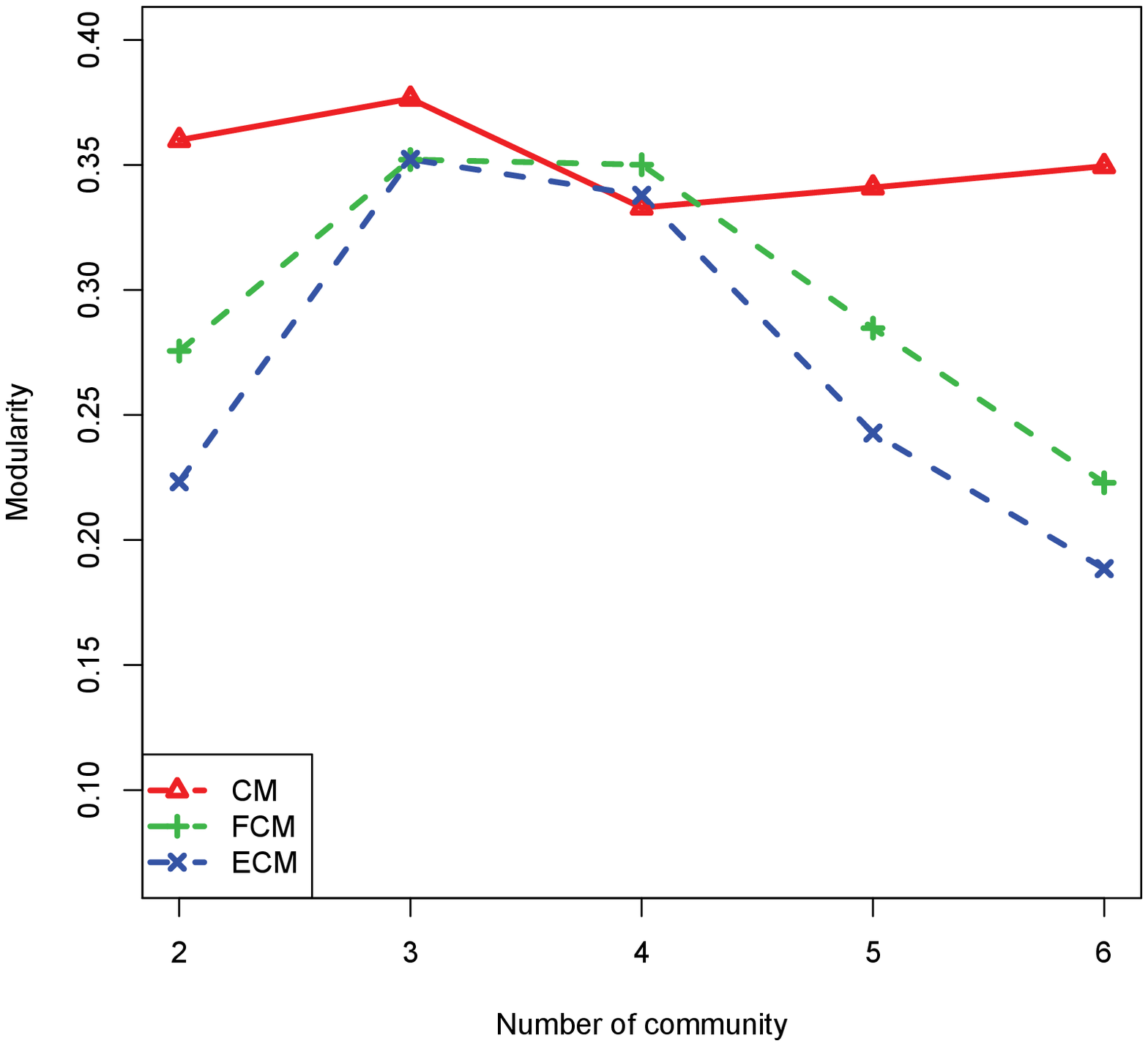}
	\hfill
	\includegraphics[width=.45\linewidth]{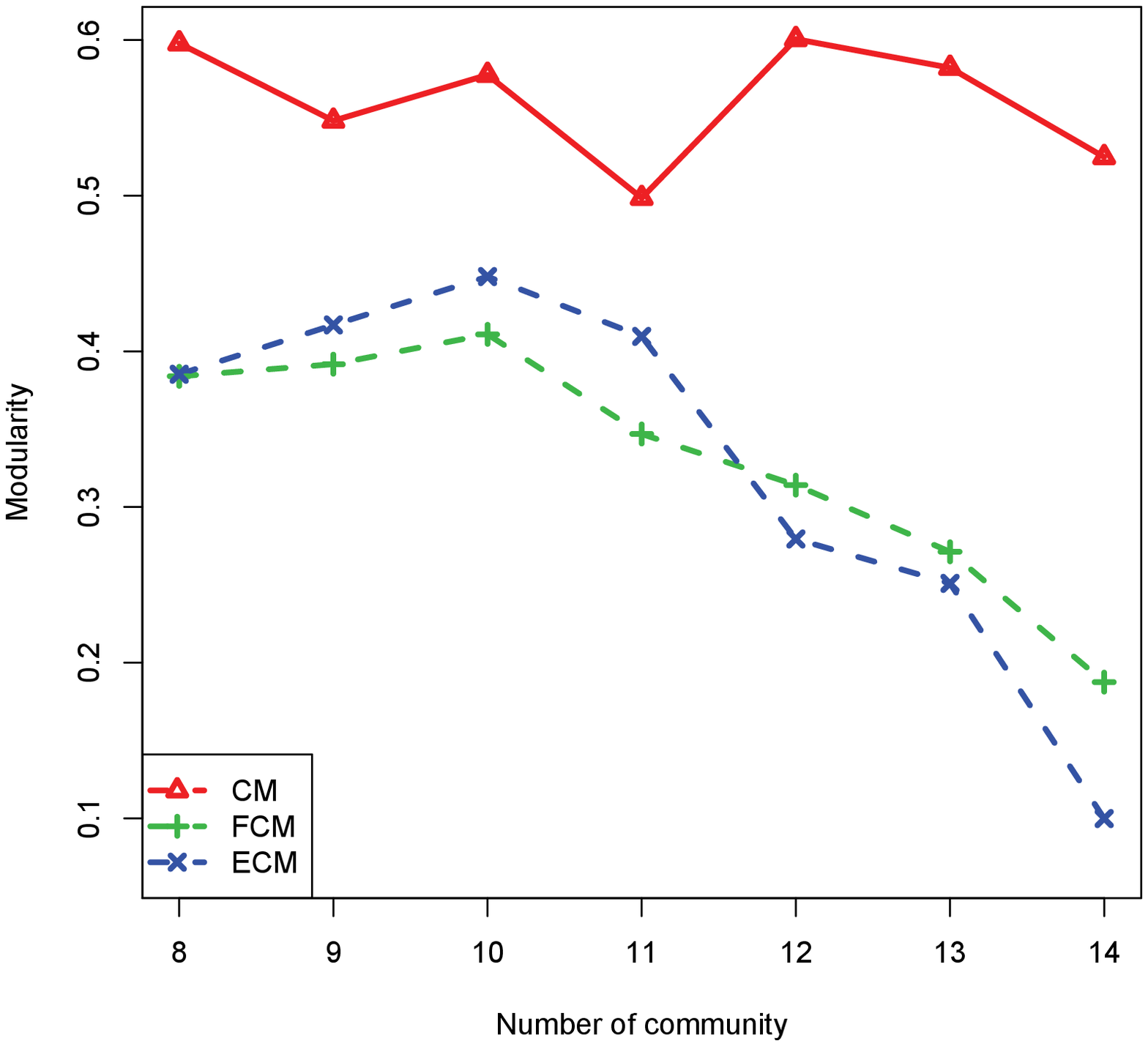}
	\hfill
    \parbox{.45\linewidth}{\centering\small a. Karate Club network}
	\hfill
	\parbox{.45\linewidth}{\centering\small b. American football network}
	\hfill
     \caption{Modularity values with community numbers.}
     \label{Qfun}
\end{figure}
\end{center}

The  modularity peaks around $c=2$ or $c=3$ as shown in Fig.~\ref{Qfun}-a. Let $c=3$, the detected communities by CM, FCM and ECM are displayed in Fig.~\ref{karate}.  As it can be seen, a
small community separated from $\omega_1$ is detected by
all the approaches.
The result by FCM shown here is got by partitioning nodes to the cluster with the highest membership. 
Zhang et al.~\cite{zhang2007identification} suggested to use a threshold $\lambda$ to covert the fuzzy membership into the final community structure. For node $i$, let the fuzzy assignment to its communities be  $\mu_{ij}, j=1,2,\cdots,c$. Node $i$ is regarded as a member of  multiple communities $\omega_{k}$ with $\mu_{ik}>\lambda$. But there is no  criterion for  determining the appropriate $\lambda$.  However, in ECM we can directly get the imprecise classes indicating our uncertainty on the actual cluster of some nodes by hard credal partitions~\cite{masson2008ecm}.

\begin{center}
\begin{figure}[!thbt]
\centering
	\includegraphics[width=0.45\linewidth]{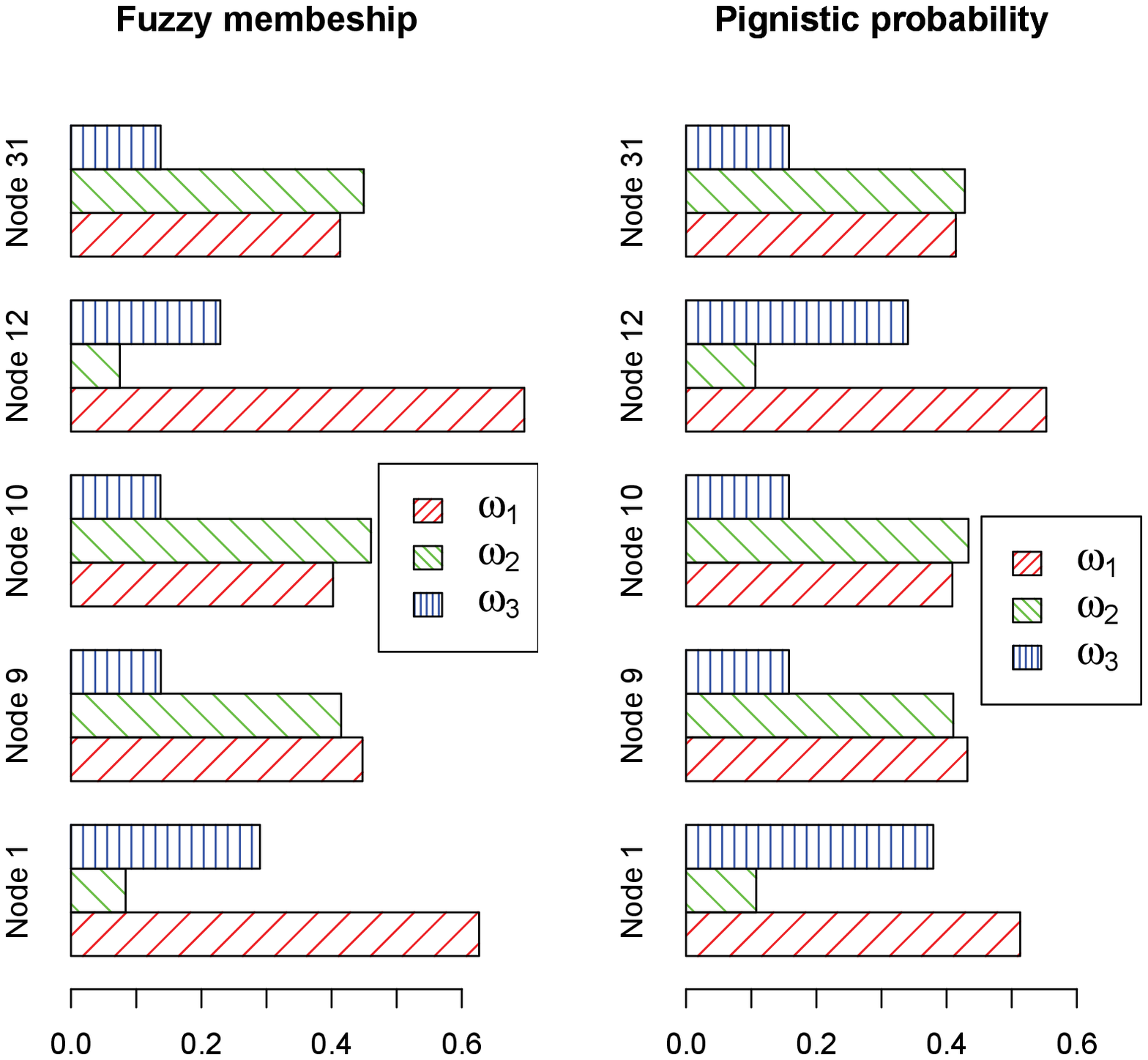}
	\hfill
	\includegraphics[width=.45\linewidth]{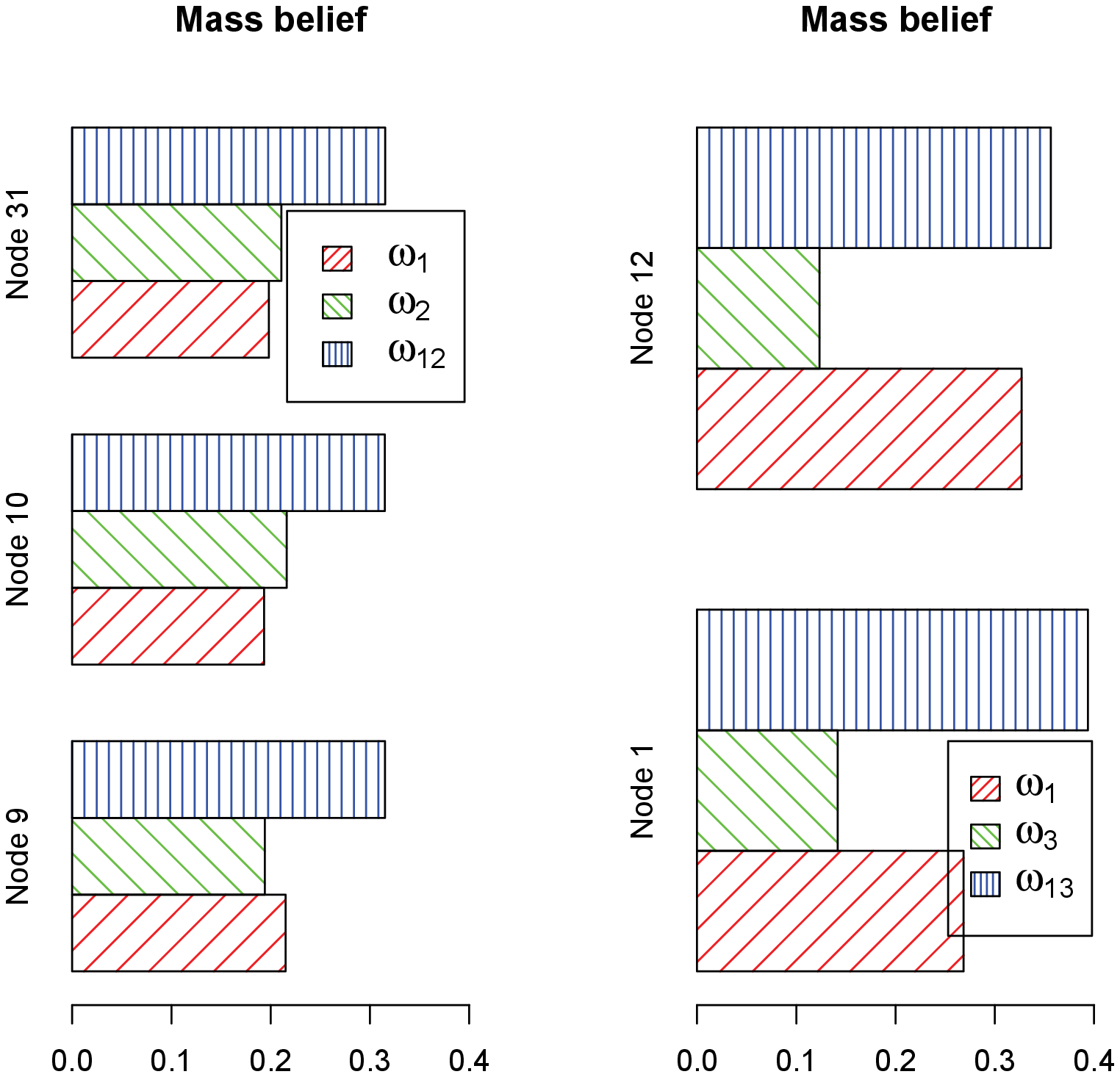}
	\hfill
    \parbox{.45\linewidth}{\centering\small a. Probability membership}
	\hfill
	\parbox{.45\linewidth}{\centering\small b. Mass belief}
	\hfill
     \caption{Clustering results of Karate club network.}
     \label{karatememb}
\end{figure}
\end{center}

As we can see in Fig.~\ref{karate}-c, for ECM, node 1,9,10,12,31 belong to two clusters at the same time. This is coincident with the conclusion in~\cite{zhang2007identification} apart from the fact that  a significant high membership value is given to $\omega_1$ for node 12 by FCM.  Actually, the case that node 12 is clustered into $\omega_{12}\triangleq \{\omega_1,\omega_2\}$ seems reasonable when the special behavior of this node is  considered. The person 12 has no contact  with others except the instructor (node 1). Therefore, the most probable class of node 12 should be the same as that of node 1. It is counterintuitive if the person 12 is partitioned into either $\omega_1$ or $\omega_2$, as it has no relation with any member in these two communities at all. The credal partition can reflect the fact that $\omega_1$ and $\omega_2$ is indistinguishable to node 12, while the fuzzy method could not. Furthermore, the mass belief assigned to imprecise classes reflects our degree of uncertainty on the clusters of the included nodes. As  illustrated in Fig.~\ref{karatememb}-b, the mass given to imprecise clusters for node 1 is larger than that to the other four nodes. This reflects our uncertain on node 1's community is largest. As node 1 is the instructor of the club, this fact seems reasonable.

Actually, the concept of credal partitions suggests different ways of summarizing data.
For example, the data can be analysed in the form of fuzzy partition thanks to the pignistic probability transformation shown in Eq.~(\ref{pignistic}). It is shown in Fig.~\ref{karatememb}-a pignistic probabilities play  the same role as fuzzy membership. A crisp partition can then be easily obtained by partitioning each
node to the community with the highest pignistic probability. In this sense, the proposed method could be regarded as a general model of hard and fuzzy community detection approaches.

\subsection{American football network}
The network we investigate in
this experiment is the world of American college football games between
Division IA colleges during regular season Fall 2000~
\cite{girvan2002community}. The vertices  in the network represent 115 teams,
while the links denote 613 regular-season games between the two teams they
connect. The teams are divided into 12 conferences  containing around 8-12
teams each and generally games are more frequent between members from the same
conference than between those from different conferences.

\begin{center}
\begin{figure}[!thbt]
\centering
	\includegraphics[width=0.45\linewidth]{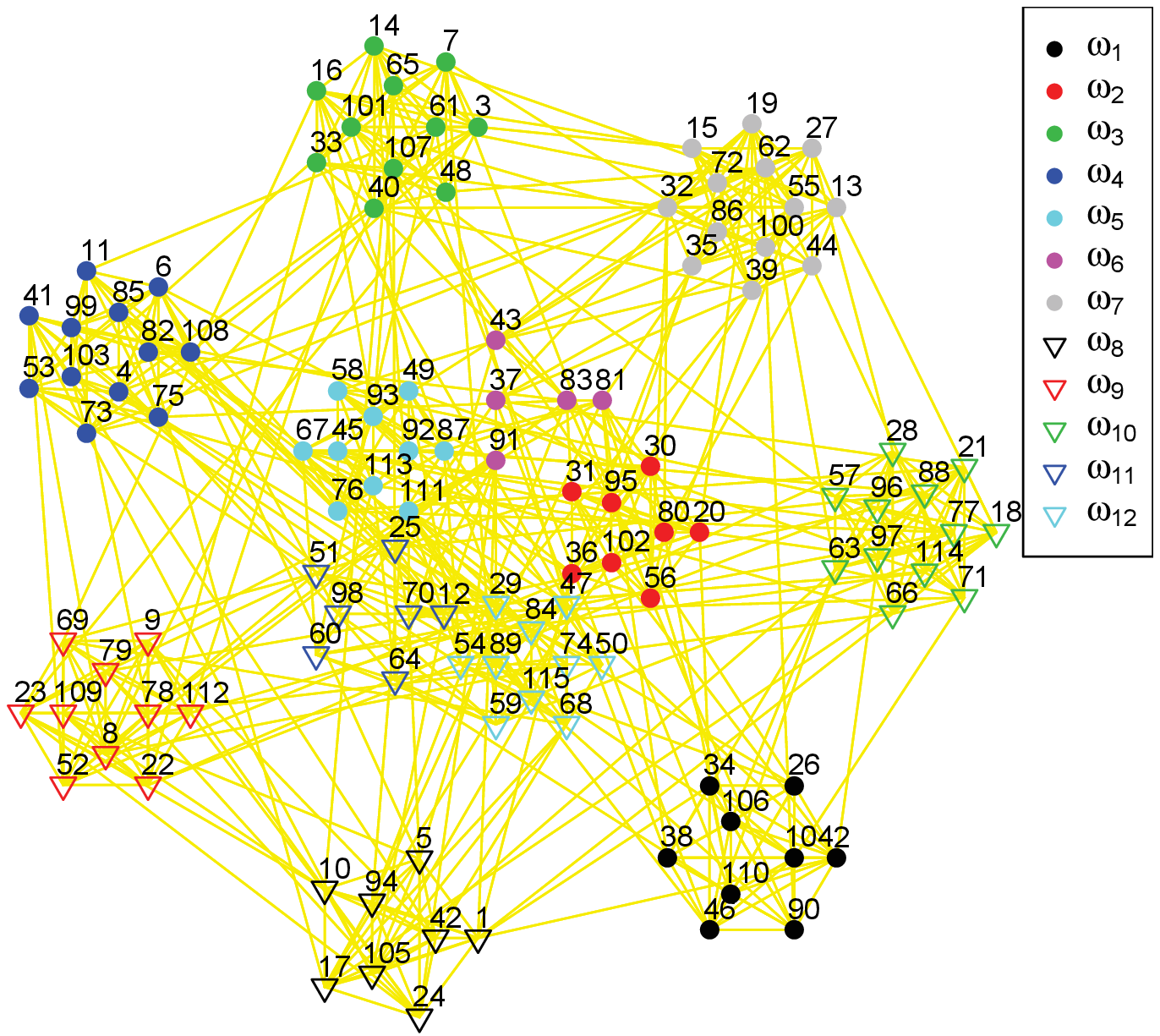}
	\hfill
	\includegraphics[width=.45\linewidth]{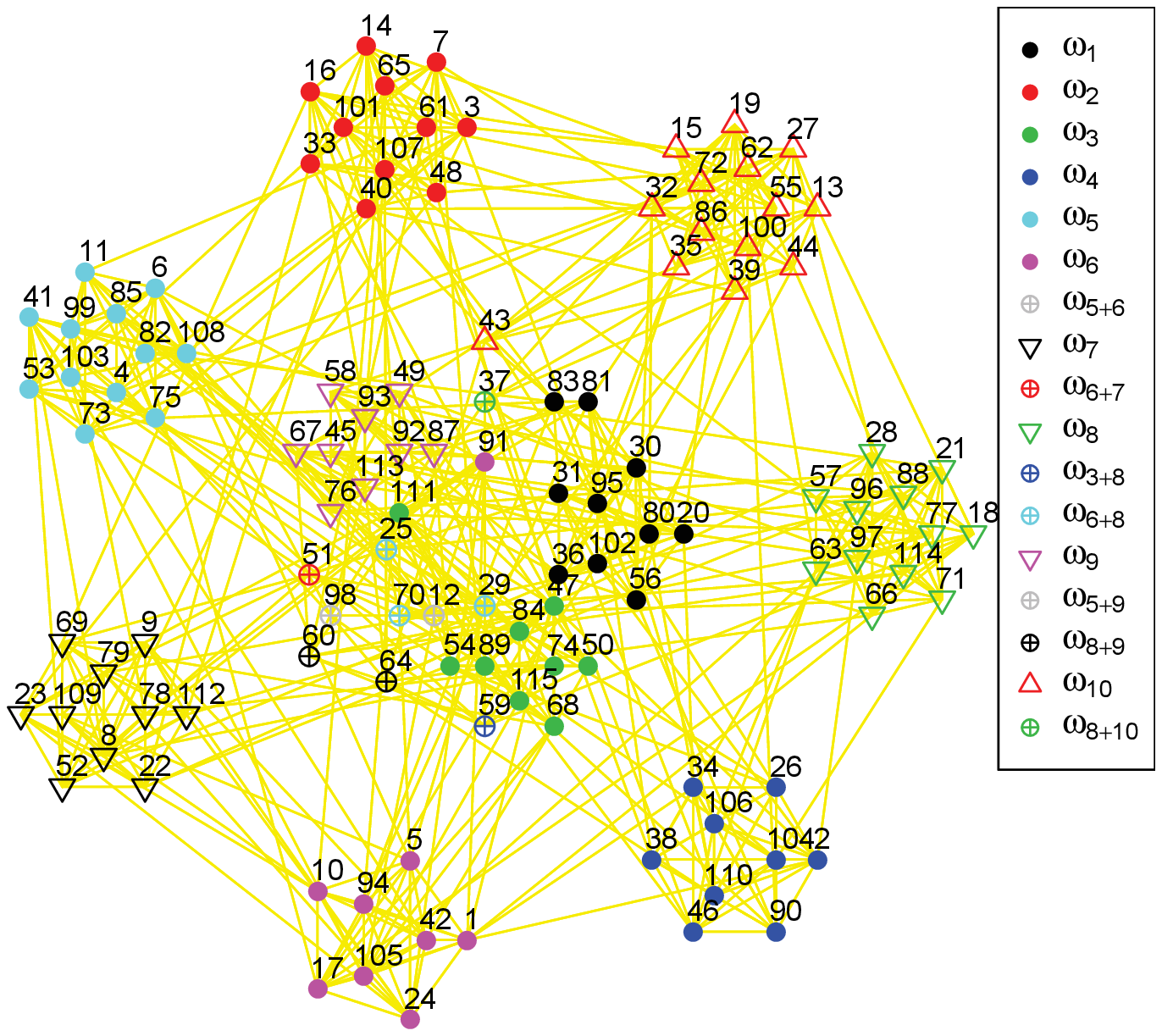}
	\hfill
    \parbox{.45\linewidth}{\centering\small a. Original network}
	\hfill
	\parbox{.45\linewidth}{\centering\small b. ECM}
	\hfill
	\includegraphics[width=0.45\linewidth]{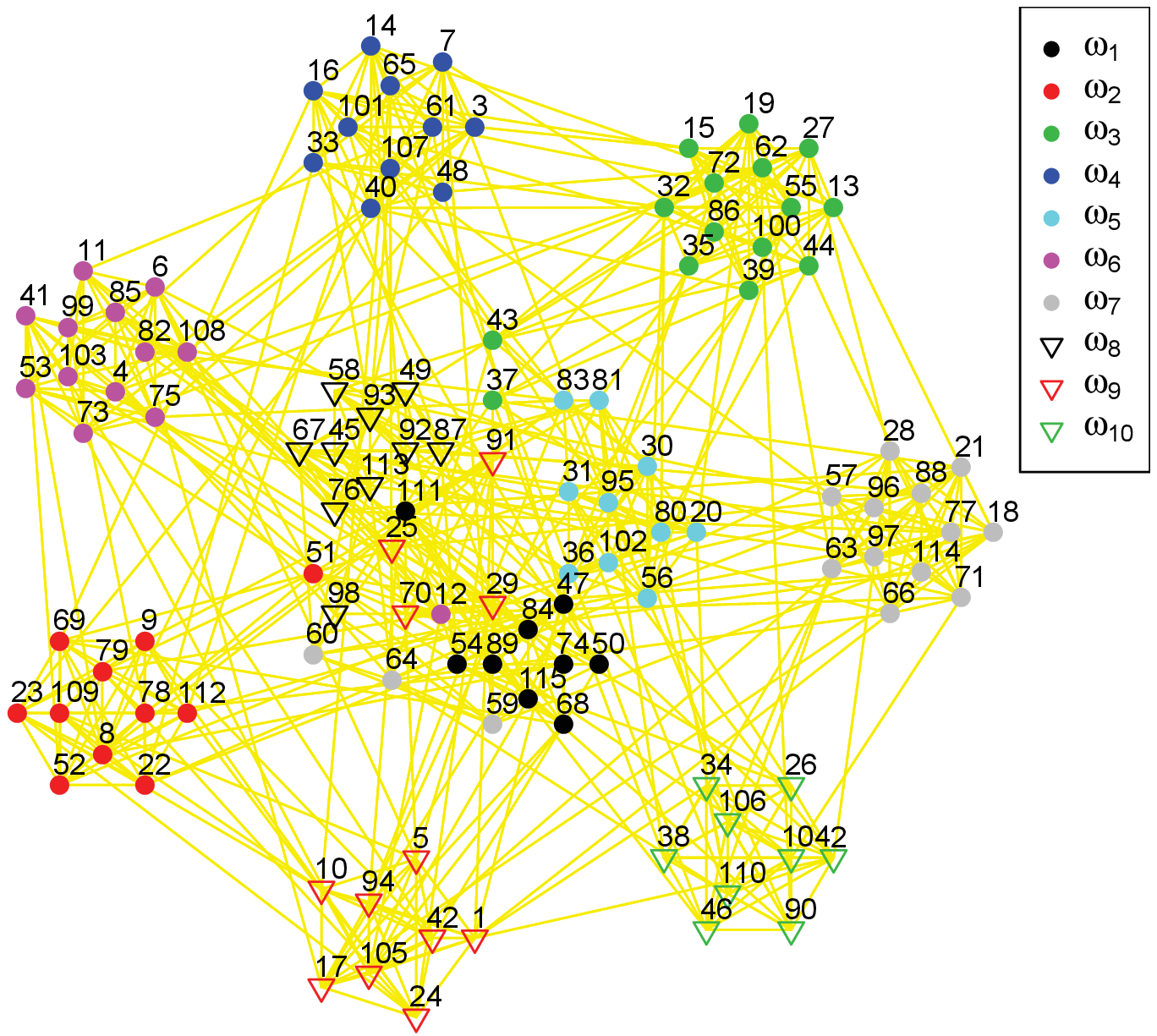}
	\hfill
	\includegraphics[width=.45\linewidth]{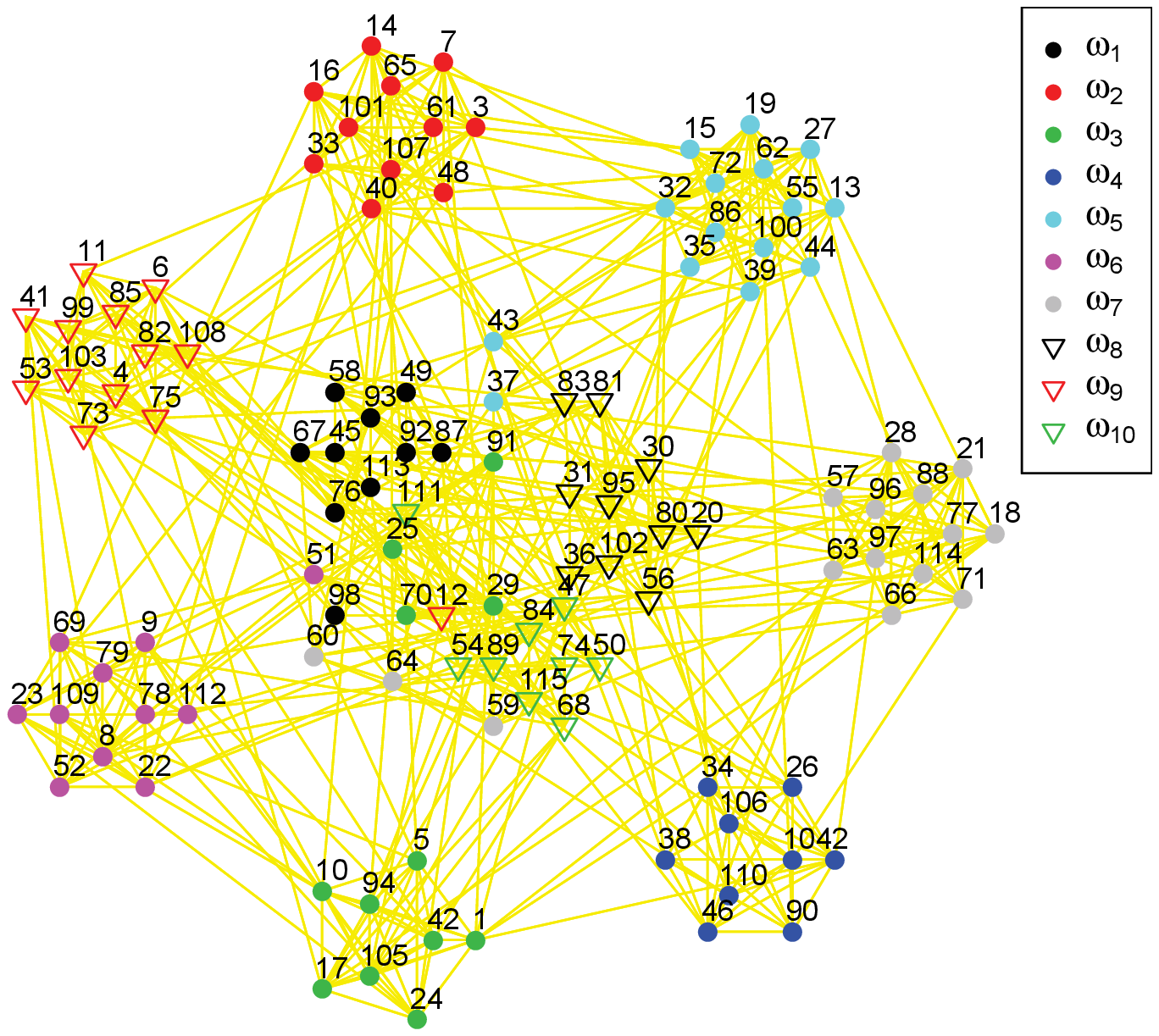}
	\hfill
    \parbox{.45\linewidth}{\centering\small c. CM}
	\hfill
	\parbox{.45\linewidth}{\centering\small d. FCM}

     \caption{American football network.}
     \label{football}
\end{figure}
\end{center}

In ECM, the number of parameters to be optimized is exponential in the
number of clusters \cite{masson2008ecm}. For the number of class larger than
10, calculations are not tractable. But we can consider only a subclass with a
limited number of focal sets \cite{masson2008ecm}. In this example, we
constrain the focal sets to be composed of at most two classes (except
$\Omega$). Fig.~\ref{Qfun}-b shows how the modularity varies with the number of communities. For credal partitions, the peak is at $c=10$. This is consensus with
the original network (shown in Fig.~\ref{football}-a) composed of 10
large communities (more than 8 members) and 2 small communities (8 members or
less than 8 members). Set $c=10$ in ECM, we can find all the ten large communities, eight of which are exactly detected. For the nodes in small communities, ECM partitions most of them into imprecise classes. As there are more than 10 communities in this
network, we use $\omega_{i+j}$ to denote the imprecise communities instead of
$\omega_{ij}$ in the figures related to this experiment to obviate
misunderstanding.

For hard partitions, nodes in small communities are simply partitioned into their  ``closest" detected cluster, which will certainly result in a loss of accuracy for the final results. Credal partitions make cautious decisions by clustering  nodes which we are uncertain into imprecise communities. The introduced imprecise clusters can avoid the risk to group a node into a specific class without strong belief. In other words, a data pair can be
clustered into the same specific group only when we are quite confident and
thus the misclassification rate will be reduced.
\vspace{-4em}
\section{Conclusion}
In this paper, a new community detection approach combing the evidential modularity, spectral mapping and evidential $c$-means is presented to identify the overlapping graph structure in complex networks. Although many overlapping community-detection algorithms have been developed before, most of them are based on fuzzy partitions. Credal partitions, in the frame of belief function theory, have many advantages compared with fuzzy ones and enable us to have a better insight into the data structure. As  shown in the experimental results for two networks in the real world, credal partitions can reflect our degree of uncertain more intuitively.  Actually, the credal partition is an extension of both hard and fuzzy ones, thus there is no doubt that more rich information of the  graph structure could be available from the detected structure by the method proposed here.  We expect that the evidential clustering approaches will be employed with promising results in the detection of overlapping communities in complex networks with practical significance.
\bibliographystyle{splncs03}
\bibliography{paperlist}
\end{document}